\documentclass[Afour,sageh,times]{sagej}

\usepackage{moreverb,url}

\usepackage[colorlinks,bookmarksopen,bookmarksnumbered,citecolor=red,urlcolor=red]{hyperref}
\usepackage[dvipsnames]{xcolor}

\newcommand\added{}
\newcommand\BibTeX{{\rmfamily B\kern-.05em \textsc{i\kern-.025em b}\kern-.08em
T\kern-.1667em\lower.7ex\hbox{E}\kern-.125emX}}

\begin{document}

\runninghead{Sanders \& Schneier}

\title{Machine Learning Featurizations for\\
AI Hacking of Political Systems}

\author{Nathan Sanders\affilnum{1} and Bruce Schneier\affilnum{2}}

\affiliation{\affilnum{1}Berkman Klein Center, Harvard University, USA\\
\affilnum{2}Belfer Center, Harvard Kennedy School, USA}


\corrauth{Nathan Sanders,
Berkman Klein Center, 
Harvard University, 
23 Everett St, \#2, 
Cambridge, Massachusetts 
02138, USA
}

\corrauth{Anonymous}

\email{nsanders@cyber.harvard.edu}

\begin{abstract}
What would the inputs be to a machine whose output is the destabilization of a robust democracy, or whose emanations could disrupt the political power of nations? In the recent essay ``The Coming AI Hackers,'' Schneier (2021) proposed a future application of artificial intelligences to discover, manipulate, and exploit vulnerabilities of social, economic, and political systems at speeds far greater than humans' ability to recognize and respond to such threats.  This work advances the concept by applying to it theory from machine learning, hypothesizing some possible ``featurization'' (input specification and transformation) frameworks for AI hacking. Focusing on the political domain, we develop graph and sequence data representations that would enable the application of a range of deep learning models to predict attributes and outcomes of political, \added{particularly legislative}, systems. We explore possible data models, datasets, predictive tasks, and actionable applications associated with each framework. We speculate about the likely practical impact and feasibility of such models, and conclude by discussing their ethical implications.
\end{abstract}

\keywords{Machine Learning, Security, Public Policy, Ethics, Featurization}

\maketitle

\section{Introduction}

\subsection{AI Hacking}

In ``The Coming AI Hackers,'' \citet{schneier_coming_2021} defines hacking as an exploitation of a system that follows its rules, but subverts its intent.  Despite the modern association of hacking with computer systems, this definition encompasses millennia of human activity: loopholes in tax law, for example.  He argues that the computerization of diverse fields, from finance to elections, increases the speed, scale, and scope of vulnerability to hacking.

With respect to the hacking of computer systems, AI is making remarkable strides.  Schneier cites several instances of specialized AI being developed and deployed to find vulnerabilities in computer code and systems automatically, enabling attackers to discover and exploit systems without human intervention \citep[p.~21]{schneier_coming_2021}.
Schneier imagines a similar AI turned to hacking social systems such as the tax code and financial regulations, or legislative and other political processes.  After all, these, like so many other systems of modern human life, are increasingly ``socio-technical systems involving computers and networks''; this leaves the social aspects of the system exposed to its technical components.

The implications of this proposal are profound in that they provoke the thought of an unknowable future where machine-generated strategies can successfully dictate outcomes of democratic political processes, and may be controlled by malicious domestic or foreign actors.
Analogizing by way of historical example, Schneier poses the question, ``Could an AI independently discover gerrymandering?''  
How about the filibuster?  
His conclusion that ``It’ll be a long time before AIs will be capable of modeling and simulating the ways that people work, individually and in groups, and before they are capable of coming up with novel ways to hack legislative processes`` raises questions: How would we get to that state?  
What approaches might AI hackers take to develop such capabilities?
What conditions would need to be satisfied for them to work?

The purpose of this paper is not to advance towards practical AI hacking as a goal, but rather to more rigorously define it \added{as it relates to political systems and, specifically, legislative processes}.  
We take the general perspective that, although there will be some benefits of the evolution of AI towards one capable of interacting competently with social systems, the advent of AI hacking as defined above would be fundamentally negative for civilization. 
Aided by a more concrete description of an AI system capable of discovering hacks of a political system, it may be possible to anticipate some of the approaches towards, and therefore ethical implications and potential dangers of, such an AI \added{such that we may mitigate their greatest risks}.

\subsection{Defining political systems}

\added{
The political system of the United States, like those of similar liberal democracies, evolves through a complex interplay of many bodies. These include institutional actors in the three branches of the federal government, state and local governments, outside interest groups aggregating the will and power of large or small constituencies, the advocates and lobbyists who represent those interests and serve as their interface to institutions, and the governed individuals who form a collective public opinion (however fragmented) expressed through their powers to vote, speak, and contribute resources. The subtle interactions between these groups and the convoluted paths by which influence is built and exerted are the primary subject of entire disciplines such as political science, public policymaking, and political communications (see e.g. \citealt{baumgartner1998basic} for a broad overview).
}

\added{
The essential premise of AI hacking for political systems is that AI may be able to approximate human-level mastery of these systems, much as it has in complex disciplines such as language arts, structural biology, and image analysis \citep{zhang2021ai}. Given sufficient data and appropriate model architectures, AI systems are quite capable of modeling long range interactions and subtle relationships between many factors like those at play in political systems. It has been specifically anticipated that lobbying firms will seek to make use of AI technologies to help automate, accelerate, and scale their capabilities \citep{nay2023large, ailobbyingnyt}.
}

\added{
While AI systems may be applicable to modeling many aspects of political systems in principle, we focus specifically on legislative processes. The expansive lawmaking powers of federal and state legislatures naturally attract immense interest among both the governed public and the monied interests that may seek to wield AI hacking tools. Moreover, the legislative processes has structure that makes it a logical first test case for AI hacking of political systems, as we will explore in more detail below. These structures include, briefly, influential communications between legislators and with the public (that are often a matter of public record) culminating in binary decisions made and recorded among a distinct set of voting members.
}

\added{
However, no interface between AI and legislative systems is possible without making a choice about how to represent the complex political processes it entails in terms of mathematical vectors and data.
Such a representation is a foundational and essential aspect of AI development.
In the following subsection, we explicate the meaning, purpose, and nature of such featurizations from the field at large before, in the following section, developing novel featurizations specifically tailored to legislative processes.
}

\subsection{Defining ML featurization}
\label{intro:feature}

Machine learning (ML) applications generally require \emph{structured} input data provided in the format of some specified ``data model'' (in the sense of, e.g., \citealt{rowe_postgres_1987}) that is tailored to the operational mechanics of the \added{ML} model \added{architecture}.  The selection of that data model is a foundational task for the application of machine learning to any domain.

There is a rich literature on the many aspects of this data model selection process, and a range of frameworks and methods that are applicable to it.
\footnote{It should be noted that the aspects described here are by no means mutually exclusive.  
A particular modeling strategy may incorporate approaches associated with several of these concepts.  Herein we cite a variety of seminal works and recent reviews to illustrate the major facets of each concept.}  
A longstanding viewpoint on data models for highly complex domains, such as human communications, is that data available in unstructured formats, such as natural language text, must be ``refined'' or ``distilled'' into more structured data suitable for algorithmic processing, namely some set of numerical vectors \citep{mccallum_information_2005}.  
The field of ``data mining'' and ``information extraction'' presents myriad techniques for this distillation for natural language and other data types \citep{balducci_unstructured_2018}.  Given input data in a format suitable for algorithmic manipulation, a primary responsibility of a machine learning developer is to do ``feature engineering'' or ``feature extraction'' \citep{khalid_survey_2014}, meaning to cull predictors from the source data that are likely to be supportive of the predictive task targeted by the model.  
Machine learning systems often rely on ``feature selection'' \citep{kira_practical_1992}, which enables models to isolate or preferentially focus on a reduced set of features that carry the greatest predictive potential.  
Generalizing this idea, the field of ``representation learning'' seeks to algorithmically construct a reduction of a complex input data format that will be optimal for some downstream predictive task or other use \citep{bengio_representation_2013}.  
``Multi-view'' models are meant to ``fuse'' data from multiple sources into a single predictive framework \citep{li_review_2016}, while ``multi-modal'' models specifically incorporate data sources with categorically different kinds of input data models (such as text and images) that may each require drastically different data representations \citep{ngiam_multimodal_2011}.  
Tools for automatic ``modality selection'' aid multi-modal modeling by identifying and privileging data modalities with the greatest predictive importance \citep{xiao_collaborative_2019}. 

Ultimately, practical systems incorporating machine learning models may be viewed as a type of ``pipeline'' facilitating the flow of input and output data between different modeling components \citep{xin_production_2021}.  
In order for this flow to proceed, the output data model from one component must match the input data model for the next, and the purpose of some components is to transform the data representation between data models.

We refer to the range of topics above in aggregate as ``featurization.''\footnote{The term of art ``featurization'' is used inconsistently. In general purpose machine learning, it is used to mean the automated process of transforming and normalizing structured variables. In bioinformatics,  it usually refers to a learned low dimensional representation of a more complex data structure. We adopt here our own somewhat expansive definition.}  We conceptualize featurization to include all steps necessary, both manual and automated, to express a complex real-world system of interest (e.g., a political process) into a mathematical format that an ML system can manipulate and operate upon.  

Prime examples of common data models and featurizations widely applied in machine learning include the following:

\begin{itemize}
    \item Images studied in computer vision, which are typically featurized as 2D or (with color information) 3D pixel arrays that can be operated on efficiently by models such as convolutional neural networks. 
    These models learn representations encoding spatial information from the input and may discover visual patterns such as the presence of a face or object.
    \item Natural language text studied in the quantitative social sciences and other fields, which is typically featurized as a token (e.g., word or character) sequence that can be operated on by models such as recurrent neural networks and transformers. 
    These models encode information about the composition and grammatical structure of a written document and may discover underlying meaning, such as references to named entities, semantic relationships, description, sentiment, or emotion.
    \item Molecules studied in cheminformatics are often represented by molecular graphs, which are composed of nodes (atoms) and edges (bonds). 
    These nodes and edges may each carry their own feature vectors describing, for example, the elemental properties of the atom and bond type. 
    These graphs can be operated on by graph neural networks that encode information about the local and global structure of the molecular graph and may discover functional groups or other substructures within the molecule that are responsible for manifesting chemical properties or bioactivity.
\end{itemize}

Specialized AI and specifically deep learning have already been applied to a variety of topics in political science, such as extracting features from political documents, measuring polarization, optimizing the geographic distribution of aid, encoding the ideology of political actors, and more \citep{chatsiou_deep_2020}. Below we explore other potential applications of AI to political processes by considering predictive tasks of potential interest to AI hackers.

\section{Frameworks for political featurization}

Here we consider possible featurizations for political systems\added{, in particular legislative processes,} that would enable predictive tasks potentially exploitable by AI hackers; specifically, graph and sequence modeling frameworks.  
In each case, we will provide a didactic description of the political system and its essential elements.
We will then frame the same elements in mathematical terms as a representation suitable for machine learning, and finally suggest predictive tasks associated with this representation.

\subsection{Graphs}
\label{feat:graphs}

Consider a network (graph) of political actors, where each node/vertex is an agent such as a person or institution and each edge represents a relationship between those actors. Edges connecting nodes could represent communication pathways between actors, such as lobbying or constituent relationships, hierarchical relations of reporting/power, or combinations of these and other relationship types.  The communication pathways may be one-way or bidirectional and may emerge or change status over time. In this conception, the manifestation of political outcomes is a consequence of communications between actors in the graph. The graphs may therefore be associated with outcomes such as the legislative disposition of a bill, the time efficiency of a process (how long it takes for legislation to move or an executive action to be taken), or the inclusion of a particular provision in a policy document.

In such a graph, the nodes are differentiated by their position in the network as well as by features such as the type of actor they represent (e.g., individual or organization), their level (e.g., position within government), their magnitude of power (e.g., seniority, budget size, constituency, etc.), and any other descriptor that may be anticipated to mediate the actor's role in the political process.  
Edges may be differentiated based on the type of relationship they represent (e.g., a constituent appeal to a representative, a lobbyist's influence on a legislator, a committee vote exercised by a member, or a backroom working relationship), the volume or frequency of communication, its age or status (e.g., current, former, or even future), and any other descriptor of the relationship's role in the political process.  
Each of these features may constitute a predictor of the outcome targeted by the model.

Nodes could even represent other entities in the political network beyond individual or organizational agents, such as issues, specific pieces of legislation, budget line items, and so on. 
Different edge types would be associated with each pair of node types; for example, the edge between a legislator and a piece of legislation could be a voting edge featurized by the legislator's current position on the legislation as well as a vector describing their voting history on the issue.

There could be many such graphs representing various parts of the political process, such as the networks of legislative relationships across a set of committees, or the networks of lobbying relationships between a legislature and a set of different interest areas.  Those graphs could carry features such as historical outcomes of the modeled process (e.g., a bill is passed or a corporation reaches a certain market cap.)

Mathematically (following, e.g., the notation of \citealt{gong_exploiting_2019} and \citealt{muzio_biological_2021}), each graph $G_k=(V, E)$ among the total number of graphs $K$ has nodes/vertices $V$, which number $n=|V|$, and edges $E$. 
Each individual edge $e_{i,j}$ connects two nodes $v_i$ and $v_j$.  The graph may be directed and weighted, in which case it can be represented by the combination of a non-symmetric adjacency tensor $A\in\mathbb{R}^{n,n,p}$, where $p$ is the number of edge features, and node feature matrix $X\in\mathbb{R}^{n,m}$, where $m$ is the number of features that describe each node.  
The graphs may have an associated vector of labels or features comprising the matrix $Y\in\mathbb{R}^{K,M}$, where $M$ is the dimensionality of the graph features. 
These symbols are visualized on a graph diagram in Figure~\ref{fig:graph}.

A variety of predictive tasks are enabled by such a representation in combination with a graph learning model such as one in the diverse class of graph neural networks (GNN) like graph convolutional neural networks and graph attention networks \citep{muzio_biological_2021}.  These tasks include:

\begin{figure*}[ht]
\vskip 0.2in
\begin{center}
\centerline{\includegraphics[width=1.3\columnwidth]{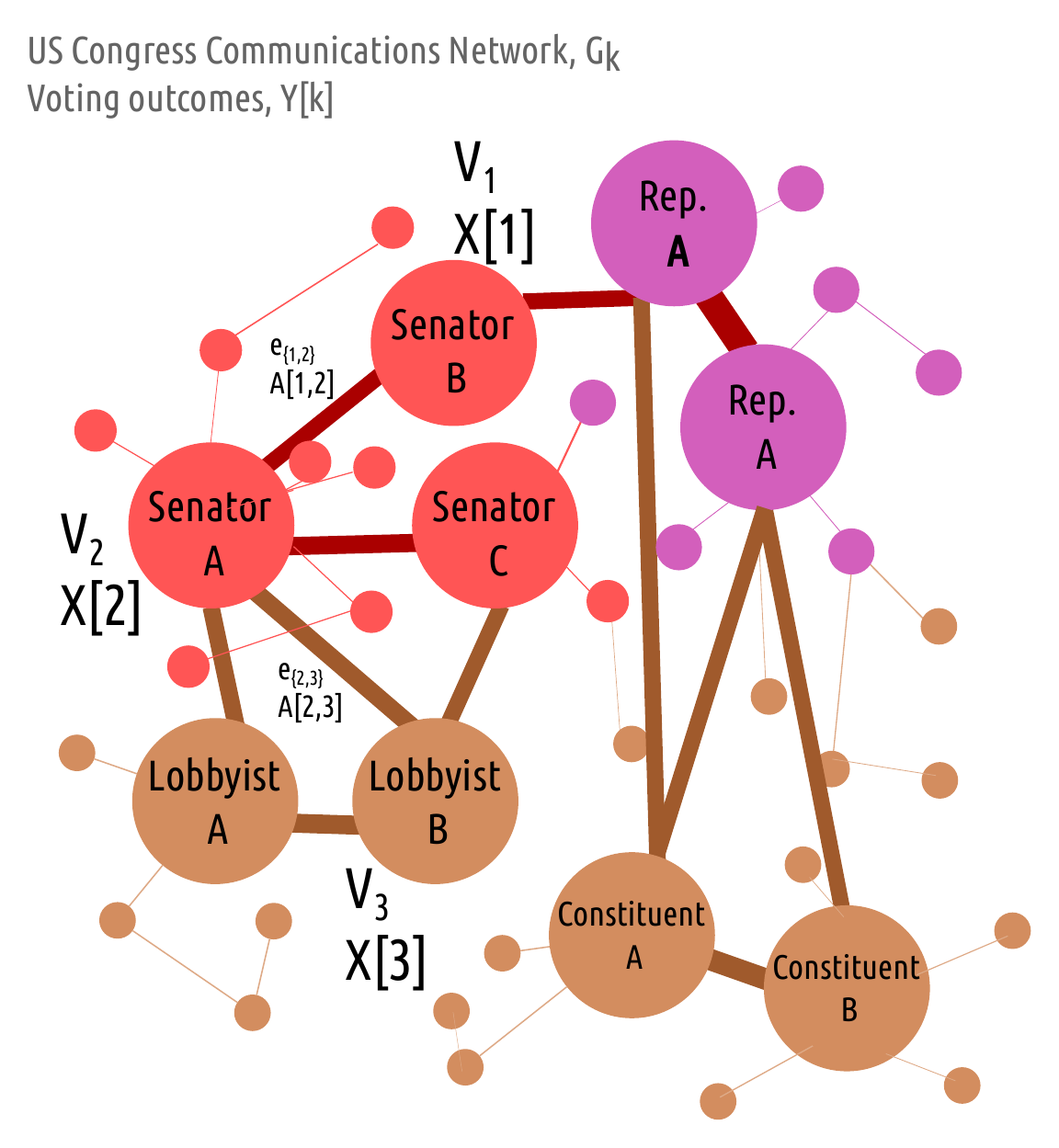}}
\caption{Illustration of a local neighborhood within a hypothetical graph representation of the US congressional legislative communication network, $G_k$. 
      The graph has a feature vector, $Y[k]$, that may represent, e.g., the body's voting outcomes across a set of bills.
      Multiple types of nodes $V$ are represented by circles, labeled as different individual members of the network: senators, representatives, lobbyists, and constituents.
      (Smaller circles represent other nodes outside of the example local network.)
      The nodes have feature vectors (e.g., $X[2]$ for Senator~A) that represent, for example, the node type (color).
      Edges (lines) connect the individuals; for example, edge $e_{\{2,3\}}$ connects Senator~A ($V_2$) to Lobbyist~B ($V_3$).
      The edge has a feature vector $A[2,3]$; for example, the width of the line may represent frequency of communication and the color may represent the type of relationship.}
\label{fig:graph}
\end{center}
\vskip -0.2in
\end{figure*}

\begin{itemize}
\item Graph label prediction (or graph classification), in which a global property (label) of a graph is predicted based on characteristics of its network structure and other metadata.  
The hacker could, for example, predict the outcome of a political process given a particular configuration of the political actor network.  
Such a predictive framework can become actionable as, for example, a search (optimization) for instantiations where the favored outcome is most likely.
For example, the model could be used to nominate a jurisdiction that may be most favorable to the introduction of legislation. 
Alternatively, a hacker could assess whether the probability of a given outcome would increase or decrease if a particular edge (communication pathway) were added to the network.  
The AI hacker could then act on this prediction by encouraging collaboration between two actors in the network.

\item Link prediction, in which the presence of an unknown edge in a network is inferred based on its local structural properties.  For example, a consistent pattern of similar actions by two political actors (nodes) with otherwise distinctive properties could imply communication (an edge) between them. A hacker targeting an inaccessible political actor could exploit this information by identifying an accessible third party actor that is discovered to be covertly in communication with the target. This could allow the AI hacker to pressure their target, without exposing their identity directly to them and without leaving any visible signature of direct communication to them.  An AI hacker could even blackmail an actor whom they can demonstrate is inappropriately communicating with another actor in the network, such as a super PAC that is unlawfully coordinating expenditures with a candidate.

\item Node attribute prediction (or classification), in which a property of a node is predicted based on its position within a network and other features.  
For example, a political actor's unstated position on an issue could be inferred based on the positions of their neighbors in the network.  
An AI hacker could gain an advantage by identifying and targeting policymakers who may be most persuadable on an issue.  
An AI hacker seeking to influence an election could also use node attribute prediction to assess the probability of a slate of potential candidates to enter an electoral race, enabling them to offer key early campaign contributions to undeclared candidates who might then become beholden to demands of the hacker.

\item Inference on node and edge feature weights or substructures, in which a model trained on historical data reveals the relative importance of each feature of its nodes and edges.  
For example, the trained weights of a fitted model for voting outcomes of a legislative body may support the inference that one factor (e.g., party alignment) is far more important than another (e.g., communication frequency) in predicting the voting behavior of each legislator. 
This insight could give an AI hacker a distinct advantage in proposing a legislative strategy.
Techniques also exist to extract explainable substructures of graphs that are associated with certain outcomes \citep{yuan_explainability_2021}. 
For example, an AI hacker might identify a pattern such as a voting block of legislators from the same region that share a particular position on a secondary issue that strongly predicts their behavior on another issue.
Such an insight could help an AI hacker to propose a communication or funding strategy targeted to that legislative block.
Moreover, this strategy is perhaps the most relevant to the charge of finding an AI system that could discover gerrymandering, which itself represents a recurring local substructure in a geographic network of constituent-district assignments.
In practice, it can be impractical to interpret or ``explain'' the complex layers of weights in deep learning models, \added{so a strategically effective pattern of this type may be difficult to detect or, in contrast, hackers may prefer to use a} predictive system that is interpretable by design \added{so they can better understand and leverage the insight surfaced by the AI} \citep{rudin_stop_2019}.

\end{itemize}

\subsection{Sequences}

Consider a sequence (an ordered list of items) of political activities, where each item is an action taken by some political actor.  Examples of actions could be steps in the legislative process for a bill, enforcement actions taken by a regulatory agency, electoral outcomes, and so on.  Each action may have some outcome associated with it, such as the size of fine issued by a regulator or the vote share in an election.

The actions in the sequence may have multivariate features that differentiate them. 
Such features may include an indicator variable for the actor who took the action, the type of action, the time it was taken, the jurisdiction of the action, the entity or topic it is related to, some measure of the magnitude of the action, background factors such as a politician's approval rating or a company's stock price, and so on.  

There are diverse machine learning methods and tasks associated with sequence modeling. Linear models such as the autoregressive integrated moving average (ARIMA) are frequently used to forecast future events based on historical sequences and their outcomes. In the deep learning domain, recurrent neural networks (RNNs) have been highly successful.  Surprisingly, convolutional neural networks, which had been more often used for image modeling and computer vision, have also proven highly effective \citep{bai_empirical_2018}.

\begin{figure*}[ht]
\vskip 0.2in
\begin{center}
\centerline{\includegraphics[width=1.0\columnwidth]{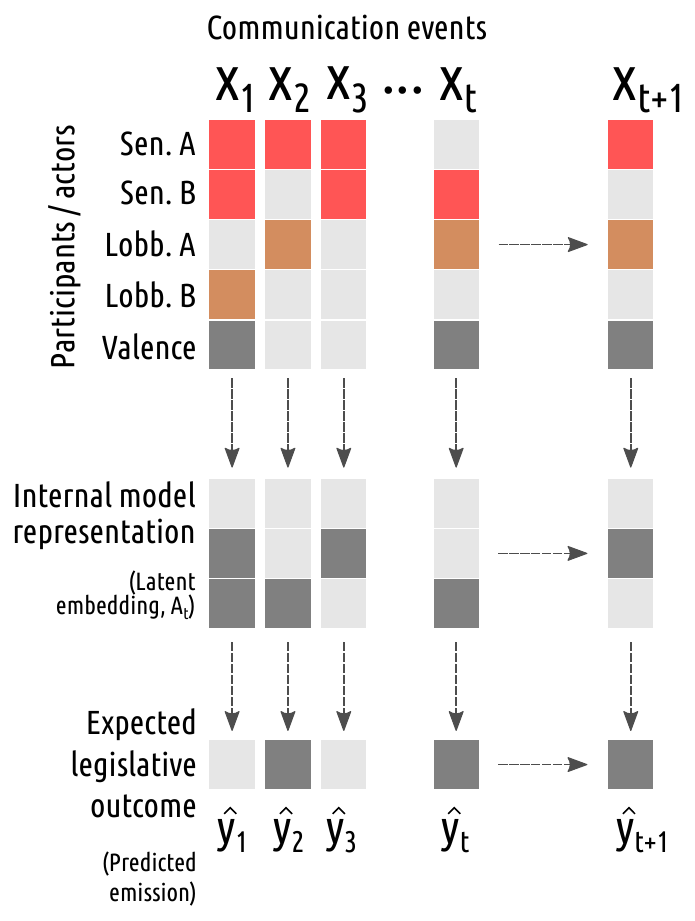}}
\caption{
      Illustration of a legislative process modeled as a sequence of communication events leading to a time-dependent legislative outcome.  
      The communication events are associated with the presence (colored blocks) or absence (grey blocks) of a set of political actors, which together comprise a binary bit vector ($x_t$).
      The sequence model translates the event bit vector and its history ($x_{<t}$) to an internal representation, the latent embedding $A_t$.
      The model then predicts an instantaneous expectation for a legislative outcome, $\hat{y}_t$, based on the latent embedding,
      The model can also extrapolate from the observed timeseries to a future communication event, $x_{t+1}$, and its associated expected outcome, $\hat{y}_{t+1}$.
      }
\label{fig:seq}
\end{center}
\vskip -0.2in
\end{figure*}

Mathematically (following the notation of, e.g., \citealt{bai_empirical_2018}), a sequence is composed of events, $x_t$, distributed over a time range, $t \in [0-T]$, each with a corresponding outcome, $y_t$. 
The variable $x$ can be multi-dimensional, carrying a set of event features, and likewise the outcome $y$ can be multivariate. 
A sequence model or ``seq2seq'' model is a mapping function, $f$, from event sequences, $x$, to predicted outcome sequences, $\hat{y}$ that is, $\hat{y}_0 \ldots \hat{y}_T = f(x_0 \ldots x_T, A_t)$.  
The tensor $A_t$ generically denotes an internal representation of the event sequence (i.e., an embedding) learned by the model.
In timeseries applications, a causality constraint is typically applied such that the inputs to $f$ for predicting $\hat{y}_t$ are limited to $x_0 \ldots x_t$, excluding any future values of $x$ at time $>t$. 
This is unnecessary for many sequence modeling applications; for example, bidirectional networks of natural language take into account both previous and subsequent textual tokens (see, e.g., \citealt{huang_bidirectional_2015} and \citealt{devlin_bert_2018}).
Such a system is illustrated in Figure~\ref{fig:seq}.

ML tasks enabled by such a representation could include the following:

\begin{itemize}
\item Supervised regression.  In this task, a sequence input is used to predict an outcome label or some other result variable.  
An AI hacker could evaluate the most likely outcome from a given sequence of events--for example, predicting the probability that a bill would be withdrawn if a particular lobbyist were to contact its lead sponsor prior to the first hearing. 
This corresponds to the generation of the outcome, $\hat{y}_t$, in Figure~\ref{fig:seq}.
\item Sequence generation.  An AI hacker could extrapolate from a series of actions by having a model generate the next action likely to be taken and its features.  
In this way, they could game out a range of likely responses to an action taken under their control, or identify the optimal sequence of events that would maximize the probability of a desired outcome. 
Moreover, a probabilistic approach to sequence generation would allow an attacker to not only weigh the probabilities of a desired outcome in any individual circumstance, but also to manage a portfolio of attacks distributed over time or in different jurisdictions to maximize their collective potential.
This corresponds to the generation of the next event bit vector, $x_{t+1}$, in Figure~\ref{fig:seq}.
\item Network inference. It is possible to infer the presence of links between political actors based on patterns in their actions, for example through point process network modeling \citep{linderman_scalable_2015,fox_contagion_2021}.  
An AI hacker might use such a technique to, for example, construct a graph of legislative communications suitable for the methods of the section on \nameref{feat:graphs} based on histories of vote or co-sponsorship sequences for a legislative body, or might uncover the most effective channels for voter persuasion around an issue based on timeseries data from social media capturing when users engaged with an issue-related hashtag.
\end{itemize}

\section{Feasibility}

Several technical factors will limit the advancement of AI hacking in the political domain.  However, in each case, we can anticipate advancements in modeling capabilities and data availability relieving those limitations over time.

First and foremost, all the predictive tasks envisioned above require the provision of labeled training data for model fitting. 
For example, training network models of the kind described above typically requires, for robust performance, hundreds of nodes for node prediction, thousands of edges for link prediction, and thousands of graphs for graph classification, and is scalable to hundreds of millions of entities \citep{hu_open_2020}. 
We know of no existing dataset that has been curated specifically for modeling the aforementioned tasks in the socio-political domain.  
However, given that there are centuries of written records of the proceedings of various political systems in diverse jurisdictions, it should be possible to construct a fairly large dataset of, for example, legislative debate and lawmaking outcomes.  
\added{Already, there is a growing practice of machine learning and natural language processing applied to legislative and other political records \citep[see e.g.][]{wilkerson_review,grimmer2022text}. For example, in recent years, researchers have distinguished passive or active bill co-sponsorship based on comparisons of bill text and legislator's speeches using graph convolutional networks and transformer models \citep{russo2022disentangling}. Others have sought to assess individual Congresspersons' attention to different policy areas based on ML classification of members' Tweets \citep{hemphill2021drives}. In another case, researchers have extracted a network of ``social'' connections between political actors and trained a graph convolutional network to learn a mathematical representation that can be applied to predictive tasks such as stance consistency \citep{feng2022political}.}

\added{Curating and extending such datasets} may require painstaking analysis of historical records to reconstruct, for example, past communication networks among a legislative body. 
Alternatively, rather than reaching back in time, an engineer building an AI hacking system could use data mining techniques to capture information about a range of contemporary political systems \citep{adnan_analytical_2019}.  
The advent of digitized communications and public records disclosures, or illicit leaks of those communications, make this scenario increasingly plausible \citep{stray_making_2019}.  
For example, a legislative communication network could be constructed from membership records with edges assigned naively based on shared committee memberships and leadership positions.  
Further, node attributes could be assigned based on party affiliation, districts, and past voting histories.  
Edge attributes could be assigned based on co-sponsorship histories.  
In jurisdictions where public hearings are routinely recorded or transcribed, characteristics of actual debate could also be featurized \citep{ruprechter_deconstructing_2020}.

Even in areas where data availability is fundamentally limited, modeling advancements may enable AI to generalize strategies learned from other datasets to successfully predict in the data-limited domain.
A robust field of research on ``transfer learning'' is concerned with exactly this problem \citep{kouw_introduction_2018}. 
In particular, the fields of ``few shot'' and ``zero shot'' learning focus on how to make predictions on tasks with extremely limited datasets \citep{xian_zero-shot_2019, wang_generalizing_2020}.  
For example, there may be instances where sufficient data exists on a modeled process, but not for a particular jurisdiction or set of political actors.  
There may be records on dozens of US states' enforcement response to emissions violations under air pollution regulations, but not yet data for a state that has newly adopted their regulatory framework.  
This may be considered a ``domain shift'' challenge and can be addressed through a variety of techniques, such as sample importance weighting \citep{wang_balanced_2017}.  
Alternatively, there may be ample data on past actions by a set of political actors, but not for the targeted task.  
For example, there may be rich historical data on the US Congress' deliberations and actions on gun control legislation, but not the relatively nascent regulatory domain of cybersecurity. 
This can be considered a ``domain adaptation'' or, more specifically, a ``concept shift'' problem.  
It too can be addressed through a variety of techniques, including finding domain-invariant feature representations or transformations, multi-task learning, and pre-training \citep{farahani_brief_2020, meftah_multi-task_2020}.  
\added{In particular, the strategy of ``fine tuning'' has been widely applied to successfully adapt pre-trained deep learning models to perform well in new, often technically complex, domains where very little training data is available \citep{gururangan2020don}, including for political tasks \citep{laurer2022less}.}

In light of all these challenges, a more viable near-term threat may be human attackers doing AI-assisted AI hacking. This would allow AI systems that are not yet fully mature to contribute to attacks in more targeted, tightly scoped ways.
For example, natural language processing (NLP) and understanding (NLU) models offer near-instantaneous analysis of copious textual documents that can be used to aid decision making. 
Particularly if applied to sensitive, private conversations (e.g. diplomatic cables leaked from the State Department or text messages harvested from hacked cell phones), such analysis could give a human political actor an unfair advantage. 

In this paper, we have focused primarily on supervised learning examples where AIs are first trained with a fixed dataset of historical examples and then applied to predict characteristics of unmeasured or hypothetical entities.  
In some cases, it may also be possible to apply reinforcement learning techniques, which explore the response surface of a reward function to learn how to optimally exploit its structure (maximize reward). 
For example, a mechanistic simulation of the political system (used as a reward function) can be used to train a reinforcement learner to take optimal actions in a real life political process.  
This methodology is analogous to the discussion of AIs learning to play the video game Breakout in \citet{schneier_coming_2021} and is similar to the use of a military war game to train combat strategists \citep[e.g.,][]{parkin_game_2020}.

\added{
Moreover, generative models may be readily exploitable by operators of these supervised learning models to capitalize on their predictions. 
Generative models, such as generative adversarial networks (GANs) and variational autoencoders, are capable of creating plausible new examples of data from the same distribution they were trained on \citep{ruthotto2021introduction}.
In the deep learning context, generative models have proven able to create remarkably human-like outputs in domains such as long and short form natural language text, photorealistic and artistic images, theoretical chemical structures, and even computer code \citep{zhang2021ai}.
A human actor looking to exploit the recommendations of a supervised AI hacking system for political processes like those described herein may operationalize their predictions by leveraging generative models to create content to be used in an influence campaign.
For example, an actor using an AI to perform link prediction to assess the persuadability of members of Congress on a policy topic could then automatically generate political messaging to distribute to each of those members using a large language model (LLM) to lobby them on this issue.
Modern LLMs may even be successful in generating precisely differentiated messaging that targets the peculiar details of the recipients policy positions, constituency, or political alignments \citep{ailobbyingnyt}.
}

\added{Finally, we note that the techniques for representation learning discussed in this work (see the section on \nameref{intro:feature}) may be applicable even to aspects of political processes whose complexity is thought to be governed by ``intangible'' or ``unmeasurable'' factors. Indeed, using data to systematically encode unstructured information emitted by complex processes into mathematical formats that can be used for downstream predictive tasks is the essential purpose of representation learning. Authors such as \cite{gagnon2020multiview,yang2021joint,feng2022political} have demonstrated the applicability of these techniques to the political domain.}

\section{Ethics, Safeguards, and Implications}
\label{sec:ethics}

\added{
We acknowledge the ethical dilemma inherent in this research; if we believe that AI hacking represents a potential threat, is it appropriate to publish a study of practical considerations for its implementation?
Will advancing general knowledge of a technology that can be exploited to achieve political ends further entrench advantaged classes in a society with a long history of applying political technologies to increase the marginalization of vulnerable groups, particular in ways that are racist, misogynist, ablist, and otherwise harmful \citep{rbenjaminbook}?
We believe the risks of AI hacking towards political systems are quite real, and we assess this dilemma in terms of the well known principle of responsible disclosure from cybersecurity \citep[see e.g.][]{herrmann2020basic}.
The introduction of AI hacking systems represents a potential zero-day exploit for democracy.
Any novel assessment of such exploits can be productively responded to with a responsible disclosure, meaning a report to the authority controlling the exploited system.
In the context of democratic political processes, the public is the controlling authority.
We have therefore pursued this work in the spirit of a responsible disclosure to the public of potential vulnerabilities in our democratic processes, and we hope this study will aid in mitigating this threat.
}

AI hacking poses a special challenge to the development of ethical AI systems. In this field, many (though certainly not all) solutions rely on regulatory engagement by the very state actors that are vulnerable to AI hacking \citep[for recent reviews, see][]{cath_governing_2018, jobin_global_2019}. 
Even in the absence of practical AI hacking, pressure for governments to take action on general-purpose machine learning has been---at best---overdue and hard-won \citep{resseguier_ai_2020}. 
The ability for an attacker to automatically disrupt legislative and regulatory action against them poses the risk of making AI hacking fundamentally ungovernable. 

\added{Indeed, when \cite{ailobbyingnyt} recently disclosed the threats to democracy potentially posed by such AI hacking systems to the general readership of the New York Times, the readership was exposed to a defensive argument generated by an AI system and published as a letter to the editor a few days later by the Times \citep{ailobbyingnytresponse}. This represents an unprecedented, but potentially harbinger, case of AI essentially defending itself against claimed threats against the political system.}

A pessimistic framing of this challenge is that of the ``Red Queen's race,'' wherein (traditionally, human) competitors engage in a continuous struggle to one-up each other's advances and, potentially, retaliate against one another \citep{taddeo_regulate_2018, asaro_what_2019, smuha_race_2021}. 
In a race to apply AI hacking tools, an aggressive party would be continuously extending their tools to overcome tactical, legal, or other barriers enacted by the defensive government or political system.  
However, if the aggressive party has unlocked the potential to automatically adjust their mode of attack in response to the actions of the defensive party, then the capacity of the latter party to escalate their defenses and keep up in the race may be short lived.  Such a scenario may reflect more of a race against time or nature rather than a race between capable competitors.  
Much like the circumstances around climate change, where policymakers face a point of no return beyond which there would be critically diminished gains from further preventative action, there may be a limited time window over which government actors can effectively forestall the impact of AI hacking on political systems.  
According to popular surveys of experts in the field, this point of no return---based on the expected performance of AI generally---could be within just a few decades \citep[e.g.][]{gruetzemacher_forecasting_2019}.

However, the future need not proceed within this pessimistic frame. 
It may be possible to structurally limit the harm potential of AI hacking systems, although the adaptability of a successful AI hacking system may make the most resilient configuration unpredictable. 
For example, distributing power across multiple institutions in a political system by providing checks and balances can limit the damage associated with AI hacking of any one lever of power, yet it would also increase the ``attack surface'' exposed \citep[as defined in cybersecurity, e.g.,][]{farrell_common-knowledge_2018, adnan_analytical_2019}.  
Similarly, it may be a viable strategy to protect sensitive functions of government by exposing them transparently to public inspection, which (in a democracy) would provide feedback to a political system that has been corrupted by an AI hacker.  
Yet recent experience in democratic politics suggests that malign actors can influence and, perhaps, corrupt public opinion through digital means \citep{lin_cyber-enabled_2019}. 
An effective AI hacker could manipulate ``common knowledge'' \citep{farrell_common-knowledge_2018} to override any outcry to their actions, even if publicly exposed.

These tradeoffs may suggest an effective strategy to control the damaging implementation of AI hacking through machine learning itself. 
A robust characterization of the performance sensitivity of practical AI hacking solutions to these tradeoffs could be generated by methods for probabilistic machine learning that help anticipate the generalization performance of models \citep[e.g.,][]{wilson_bayesian_2020}. 
Such an analysis could determine what instantiations of a featurized political system would be least vulnerable to an AI hacker. This sensitivity surface could then be optimized to identify a political configuration that minimizes risk. 
Such an optimization would require complete knowledge of, or access to, the adversarial AI hacking algorithm, or at least a structurally similar one.  
Perversely, the best defense against an AI-algorithm hacker may be another, white hat defensive AI algorithm that can simulate and assess shortcomings in the attacking algorithm.

Another safeguard against AI hacking may be the inherent difficulty in hacking political systems, regardless of the sophistication of the machine learner.  After all, reliably achieving political outcomes is a task that generations of humanity's own most well-meaning and intelligent actors---as well as malignant and/or less intelligent actors---have failed at. There are many tasks at which modern machine learning systems simply fail to perform. Worse, there are many tasks that ML systems may appear to solve, yet will actually fail to generalize to more complex or realistic examples \citep{damour_underspecification_2020, geirhos_shortcut_2020}.

A tool to recognize when a policy has been manipulated could be a further safeguard against AI hacking.  Likewise, the advent of ``deepfakes'' (hyperrealistic computer-generated audio and video) has spurred development of fake-spotting systems and models \citep{wang_fakespotter_2020}. Notwithstanding the potential for a sufficiently advanced AI to fool the spotting system, the need for such techniques could again motivate the systematic study of AI hacking by benign researchers \added{\citep{crothers2022machine}}.

We note \added{that there is significant} structural inequity in the challenge posed by AI hacking to democratic systems. If a polity fears that policy changes may have been dictated by a manipulative AI system, they may be inclined to resist change and to introduce additional friction into the policymaking process. This may indeed be a valid mitigating factor against AI hacking.  But, in this way, fear of AI hacking may promote conservative modes of governing that are skeptical of progressive change. The legitimate risks associated with practical applications of AI hacking in the present day, and their growth over time, should be carefully considered in any systemic response.

\added{
Ultimately, coordination between machine learning and social scientists is critical to minimize the sociotechnical risks posed by AI technologies \citep{andrus2020ai}. 
Interdisciplinary collaborations should engage in thinking through the potential uses, feasible designs, and risks associated with these technologies so that safeguards like those described here may be evaluated and implemented. 
}

\section{Conclusion}

\added{We have considered the practical implications of the forecast advent of ``AI hackers'' for political, particularly legislative, processes. 
We have explored what potential implementations of AI systems for hacking legislative processes may looklike, developing featurization (input specification and transformation) frameworks that would allow two existing, well developed classes of machine learning models (graph learners and sequence learners) to be applied to tasks of legislative politics.
For each modeling approach, we discussed potential datasets and specific predictive tasks which would enable a human directing an AI hacker to take advantage of such capabilities, while also discussing their feasibility, limitations, and ethical considerations.
}

\added{
We hope this and other work exploring the potential applications of ``AI hacking'' will help social scientists, policymakers, and ML researchers to anticipate and mitigate the potential risks and threats of AI to political processes. 
If we are to be thrust into a ``Red Queen's race'' parrying between offensive and defensive developments in AI hacking, we should be proactive in developing and updating threat models, detection techniques, regulatory frameworks, and other mitigating measures to protect our democratic systems.
}

\bibliographystyle{SageH}
\bibliography{references.bib}

\begin{acks}
We thank Rebecca Tabasky and the Berkman Klein Center for Internet and Society for  facilitating conversations about this topic at the May 2021 Festival of Ideas event. We thank the editor and anonymous reviewers for helpful suggestions.
\end{acks}

\end{document}